\documentclass[dvips]{elsarticle}
\usepackage{amsmath,amsfonts,graphicx,amssymb,array,url}
\begin{document}
\title{Exact transparent boundary conditions for the parabolic wave equations with linear and quadratic potentials}
\author[lpi]{R.M.~Feshchenko\corref{cor1}}
\ead{rusl@sci.lebedev.ru}
\author[izmiran]{A.V.~Popov}
\ead{popov@izmiran.ru}
\address[lpi]{P.N. Lebedev Physical Institute of RAS, 53 Leninski Pr., Moscow, Russia, 119991}
\address[izmiran]{Pushkov Institute of Terrestrial Magnetism, Ionosphere and Radiowave Propagation of RAS, Troitsk, Moscow region, Russia, 108840}
\cortext[cor1]{Corresponding author}
\begin{abstract}
In this paper exact 1D transparent boundary conditions (TBC) for the 2D parabolic wave equation with a linear or a quadratic dependence of the dielectric permittivity on the transversal coordinate are reported. Unlike the previously derived TBCs they contain only elementary functions. The obtained boundary conditions can be used to numerically solve the 2D parabolic equation describing the propagation of light in weakly bent optical waveguides and fibers including waveguides with variable curvature. They also are useful when solving the equivalent 1D Schr\"odinger equation with a potential depending linearly or quadratically on the coordinate. The prospects and problems of discretization of the derived transparent boundary conditions are discussed.
\end{abstract}

\maketitle

\section{Introduction} 
Leontovitch and Fock introduced the parabolic wave equation (PWE) (also known as the Fresnel equation) about fifty years ago \cite{fock1965electromagnetic}. The PWE is widely used in radiophysics and oceanic acoustics for modeling electromagnetic wave and sound propagation \cite{tappert1977parabolic, levy2000parabolic, papadakis1994exact}. An important application of PWE has been found in visible light and X-ray optics, where PWE is used to describe the propagation of weakly divergent light beams in inhomogeneous media \cite{OptCom_Kopylov_1995}. The PWE can be also used to describe coherent scattering phenomena in X-ray imaging \cite{artyukov2014coherent}.

A lot of papers have been dedicated to the refinement of the parabolic equation method in order to increase its numerical accuracy and adjust it to specific physical problems (wide-angle version \cite{claerbout1976fundamentals}, vectorial PWE \cite{popov2000modeling}, etc). However, there are important applications where the original Leontovich's approximation still provides excellent accuracy and numerical efficiency.  In particular, in hard X-ray optics, where optical constants are close to unity, Leontovich’s PWE is a powerful tool for modeling complex optical structures such as nano-waveguides \cite{jarre2007x} and diffraction zone plates \cite{OptCom_Kopylov_1995, attwood2007soft}.

No matter what kind of parabolic equation is concerned, any numerical solution, for instance, by a finite difference (FD) approximation, requires appropriate boundary conditions (BC) as realistic computational domains are necessary finite, while the original wave field might be sought in the whole space. Such a BC must substitute accurate wave field calculations outside the finite computational domain with some relations between its boundary values \cite{ComCompPhys_Antoine_2008}.

There are two main approaches to the computational domain truncation for the parabolic type equations. The first one is based on an exact analytical solution of the governing PE in the outer domain, free of diffracting objects. Projection of such a solution onto the computational domain border leads to so called transparent boundary conditions (TBC) \cite{WM_Baskakov_1991, RadSci_Popov_1996}.  Completely different idea underlies the perfectly matched layer (PML) techniques  \cite{JCompPhys_Berenger_1994,yevick1997analytic}: the assumption that a gradual change of the medium parameters by adding small absorption will cause the outgoing wave attenuation without producing backward reflection. Both approaches have their advantages and shortcomings: the former one is in a sense exact but may involve sophisticated derivations whereas the letter is analytically simpler but basically of approximate nature. In this paper we will only deal with the former type, which is based on assumption that any wave that reaches the boundary from within of the computational domain propagates outward and never returns. TBCs are generally non-local Neumann-to-Direchlet or Direchlet-to-Neumann mappings relating the wave field amplitude boundary values with its first derivative a coordinate. 

For a general review of TBCs and their applications see \cite{ComCompPhys_Antoine_2008}. Here we confine ourselves with the original Leontovich 2D PWE. For this equation, i.e. Eq. (\ref{0a}) (see below) with $\alpha=0$, a TBC was formulated about twenty years ago and now is known as Basakov-Popov-Papadakis (BPP) condition \cite{WM_Baskakov_1991,papadakis1994exact,RadSci_Popov_1996,JCompPhys_Yevick_2001}. A TBC for the linear potential (see Eq. (\ref{0b}) below) but for constant coefficient $g(z)=1$ has also been known for a long time \cite{RadSci_Popov_1996}, although, involving integration of a ratio of Airy functions to obtain the kernel, it is quite complicated and poses difficulties for the numerical implementation \cite{ehrhardt2004solutions}. In this paper, extending our previous work \cite{feshchenko2014parabolic} we contrive to obtain two TBCs for the 2D PWE: (i) one for the linear potential with varying coefficient $g(z)$ (ii) and one for  the quadratic potential (see Eq. (\ref{0c}) below). These TBCs do not involve special functions and additional integration, having kernels explicitly expressed via elementary functions. We also try to preserve as much generality as possible by considering in the linear potential case the curvature $g(z)$ to be an arbitrary positive function of $z$. 

\section{Methods}
In this paper we will only be concerned with the linear 2D PWE having the following form
\begin{equation}
2ik\frac{\partial u}{\partial z}+\frac{\partial^2u}{\partial x^2}+k^2\alpha(x,z)u=0,
\label{0a}
\end{equation}
where $x$ and $z$ are coordinates, $k=2\pi/\lambda$ is the wave number and $\alpha(x,z)$ is a finite function. The computational domain is defined as $-x_0<x<x_0$ where $x_0$ is a positive number. We will also assume that $u(0,x)=0$ when $|x|>x_0$. The Eq. (\ref{0a}) is a full analog of the 1D Schr\"odinger equation (SE), where $z$ is replaced with time and $\alpha$ is the potential. 

The function $\alpha(x,z)$ generally comprises both on the dielectric permittivity of the medium and on a particular choice of the coordinate system \cite{popovdiffraction}. For instance, in a weakly bent optical waveguide or fiber we can introduce a curvilinear system of coordinates so that the propagation of light is described by the ordinary PWE but with a fictitious dielectric permittivity, which is a sum of the true dielectric permittivity $\alpha_0(x,z)$ and an additional term resulting from this particular choice of coordinates. 

We will consider two cases for the dependance of $\alpha(x,z)$ on the transversal coordinate $x$. The first case is that of the linear dependance with the coefficient itself dependent on $z$ as is shown in the following expression 
\begin{equation}
\alpha(x,z)=\alpha_0(x,z)+ag(z)x,
\label{0b}
\end{equation}
where $\alpha_0(x,z)$ is a function having a compact support, i.e. $\alpha_0(x,z)=0$ outside the the computational domain when $|x|>x_0$, $a$ is a constant and $g(z)$ is a real positive function. So, outside the computational domain we have the PWE with linear dependence of $\alpha(x,z)$ on coordinate $x$. It can be shown that in the case of weak bending $g(z)=2/R(z)$ in the second term of Eq. (\ref{0b}) . In other words $g(z)$ is the curvature of the waveguide or fiber and $R(z)$ is the curvature radius.

In the second case function $\alpha(x,z)$ can be expanded further up to the second order by coordinate $x$ as
\begin{equation}
\alpha(x,z)=\alpha_0(x,z)+bx^2,
\label{0c}
\end{equation}
where $\alpha_0(x,z)$ has a compact support as in the previous case and $b$ is a parameter. In (\ref{0c}) we omitted the linear term as it can be always eliminated by a simple shift of the coordinate system. The 2D PWE with the quadratic dependence of the potential term on a coordinate may have applications in the waveguide theory in optics and acoustics. In case of the quantum mechanics and 1D SE the quadratic potential describes a particle in a parabolic trap (in case the potential is attractive), which is itself an interesting problem, and which numerical solution can be greatly facilitated by an appropriate TBC.

We will derive exact TBC from Eq. (\ref{0a}) with both linear and quadratic dependence of $\alpha$ on $x$ by applying to it the Laplace transform by coordinate $x$. The transformed equations will be linear and of the first order relative the partial derivatives and susceptible to the solution by standard methods of the mathematical physics \cite{miersemann2014partial}. The final result is obtained by applying the reverse Laplace transform to the obtained solutions.

\section{TBC for the linear potential}
Let us consider the PWE (\ref{0a}) with function $\alpha$ from (\ref{0b}). Then beyond and at the boundary of the computational domain defined above as $|x|<x_0$, where $x_0$ is a positive number, it takes the following form
\begin{equation}
2ik\frac{\partial u}{\partial z}+\frac{\partial^2u}{\partial x^2}+k^2axg(z)u=0,
\label{1a}
\end{equation}
where the parameters have been defined before. It can be transformed by introducing a new variable $\varphi(x,z)$:
\begin{equation}
u(x,z)=\varphi(x,z)\exp\left[i\frac{akx}{2}G(z)-i\frac{ka^2}{8}\int\limits_0^z G^2(\mu)\,d\mu\right],
\label{1b}
\end{equation}
where the function $G(z)$ is defined as
\begin{equation}
G(z)=\int\limits_0^z g(\nu)\,d\nu.
\label{1c}
\end{equation}
By substituting (\ref{1b}) into (\ref{1a}) one obtains an equivalent equation for the introduced function $\varphi$:
\begin{equation}
2ik\frac{\partial \varphi}{\partial z}+\frac{\partial^2\varphi}{\partial x^2}+ikaG(z)\frac{\partial \varphi}{\partial x}=0.
\label{1d}
\end{equation}
The advantage of this equation is independence of its coefficients of coordinate $x$, which allows us, for instance, to obtain a TBC for (\ref{1d}) by applying the Laplace transform by variable $x$. 

The Laplace transform of function $\varphi$ is defined as
\begin{equation}
F(w,z)=\int\limits_0^\infty\varphi(x,z)\exp(-wx)dx.
\label{1e}
\end{equation}
After applying to Eq. (\ref{1d}) the following expression is obtained
\begin{equation}
\frac{\partial F}{\partial z}=\left[\frac{iw^2}{2k}-\frac{aw}{2}G(z)\right]F-\left[\frac{iw}{2k}-\frac{a}{2}G(z)\right]\varphi_0(z)-\frac{i}{2k}\varphi'_0(z),
\label{1f}
\end{equation}
where
$$
\varphi_0(z)=\varphi(0,z),\quad \varphi'_0(z)=\frac{\partial \varphi(0,z)}{\partial x}.
$$
If we intend to derive a TBC simulating open upper half-space, we shall assume that the initial wave field $u(x,0)=0$ for $x>0$ -- the sources of the wave field are located where $x<0$. This means that $F(w,0)=0$ too. Now a unique solution of Eq. (\ref{1f}) can be written as
\begin{multline}
F(w,z)=\\
-\int\limits_0^z\exp\left[i\frac{w^2}{2k}(z-\zeta)-\frac{aw}{2}\int\limits_\zeta^z G(\mu)d\mu\right]\left\{\left(\frac{iw}{2k}-\frac{a}{2}G(\zeta)\right)\varphi_0(\zeta)+\frac{i}{2k}\varphi'_0(\zeta)\right\}d\zeta.
\label{1g}
\end{multline}
The reverse Laplace transform of $F(w,z)$ is
\begin{equation}
\varphi(x,z)=\frac{1}{2\pi i}\int\limits_{c-i\infty}^{c+i\infty}F(w,z)\exp(wx)dw.
\label{1h}
\end{equation}
In expression (\ref{1h}) there is an undefined parameter $c$, which is chosen so that all singularities of the Laplace image $F$ lie to the left from the integration path. Since the sources of the wave field are located where $x<0$, the field should decrease when $x\to\infty$. So, it follows from (\ref{1e}) that the function $F$ has no poles to the right of the axis $\mathrm{Im}(w)=0$. Therefore, parameter $c$ can be chosen equal to zero.

Now applying back transform (\ref{1h}) to expression (\ref{1g}) and taking into account that
\begin{align}
\frac{1}{2\pi i}\lim_{x\to0}&\int\limits_{-i\infty}^{i\infty}F(w,z)\exp(xw)dw\label{1i}=\varphi_0(z),\\
\frac{1}{2\pi i}\lim_{x\to0}&\int\limits_{-i\infty}^{i\infty}F(w,z)\exp(xw)wdw\label{1j}=\varphi'_0(z),\\
\lim_{x\to0}&\int\limits_{-i\infty}^{i\infty}\exp(i\beta w^2)\exp(xw)dw=\sqrt{\frac{\pi i}{\beta}}\label{1k},\\
\lim_{x\to0}&\int\limits_{-i\infty}^{i\infty}\exp(i\beta w^2)\exp(xw)wdw=-2\pi\delta(\beta)\label{1l},\\
\lim_{x\to0}&\int\limits_{-i\infty}^{i\infty}\int\limits_0^z f(\zeta)\exp(i(z-\zeta)w^2/2k)\exp(xw)w^2d\zeta dw\notag	\\
&=\sqrt{\frac{\pi}{i}}(2k)^{3/2}\frac{\partial}{\partial z}\int\limits_0^z\frac{f(\zeta)}{\sqrt{z-\zeta}}d\zeta\label{1m}.
\end{align}
we can write down the following two equivalent TBCs
\begin{multline}
\varphi_0(z)=-\frac{1}{\pi\sigma}\int\limits_0^z\exp\left[i\frac{ka^2}{8}(z-\zeta)\overline{G}^2\right]\frac{\varphi'_0(\zeta)}{\sqrt{z-\zeta}}d\zeta\\
+i\frac{ka}{2\pi\sigma}\int\limits_0^z\exp\left[i\frac{ka^2}{8}(z-\zeta)\overline{G}^2\right]\frac{\varphi_0(\zeta)(\overline{G}-2G(\zeta))}{\sqrt{z-\zeta}}d\zeta,
\label{1o}
\end{multline}
\begin{multline}
\varphi'_0(z)=-\sigma\frac{\partial}{\partial z}\int\limits_0^z\exp\left[i\frac{ka^2}{8}(z-\zeta)\overline{G}^2\right]\frac{\varphi_0(\zeta)}{\sqrt{z-\zeta}}d\zeta\\
+i\sigma\frac{ka^2}{4}\int\limits_0^z\exp\left[i\frac{ka^2}{8}(z-\zeta)\overline{G}^2\right]\frac{\varphi_0(\zeta)(G(z)-G(\zeta))\overline{G}}{\sqrt{z-\zeta}}d\zeta\\
-\sigma\frac{a}{4}\int\limits_0^z\exp\left[i\frac{ka^2}{8}(z-\zeta)\overline{G}^2\right]\frac{\varphi'_0(\zeta)\overline{G}}{\sqrt{z-\zeta}}d\zeta.
\label{1p}
\end{multline}
where
$$
\overline{G}=\frac{1}{z-\zeta}\int\limits_\zeta^z G(\mu)d\mu,\quad \sigma=\sqrt{\frac{2k}{\pi i}}.
$$
We shall note here that expressions (\ref{1i})--(\ref{1k}) are quite obvious, but for the derivation of Eq. (\ref{1l}) and (\ref{1m}) see appendix \ref{ap1}. The expression (\ref{1l}) is especially important as it causes doubling of the coefficients in front of right hand sides of Eq. (\ref{1o}) and (\ref{1p}).

Expressions (\ref{1o}) and (\ref{1p}) are Newman-to-Dirichlet mappings that relate function $\varphi$ or its transverse derivative at $x=0$ to the boundary values of the solution and its derivative at all previous positions between zero and the current position at $z$. Furthermore, because Eq. (\ref{1d}) does not depend explicitly on $x$, the obtained conditions are valid for any $x=x_0$.

Using definition (\ref{1b}), we can  obtain the boundary conditions for the wave field amplitude $u$ itself. The final equations are (for $x=x_0$)
\begin{multline}
u(x_0,z)=-\frac{1}{\pi\sigma}\int\limits_0^z\exp[i\Phi(z,\zeta)]\frac{u'_x(x_0, \zeta)}{\sqrt{z-\zeta}}d\zeta\\
+i\frac{ka}{2\pi\sigma}\int\limits_0^z\exp[i\Phi(z,\zeta)]\frac{u(x_0,\zeta)(\overline{G}-G(\zeta))}{\sqrt{z-\zeta}}d\zeta,
\label{1q}
\end{multline}

\begin{multline}
u'_x(x_0,z)+\sigma\frac{a}{4}\int\limits_0^z\exp[i\Phi(z,\zeta)]\frac{u'_x(x_0,\zeta)}{\sqrt{z-\zeta}}(\overline{G}-G(z))d\zeta=\\
-\sigma\frac{\partial}{\partial z}\int\limits_0^z\exp[i\Phi(z,\zeta)]\frac{u(x_0,\zeta)}{\sqrt{z-\zeta}}d\zeta\\
+i\sigma\frac{ka^2}{8}\int\limits_0^z\exp[i\Phi(z,\zeta)]\frac{u(x_0,\zeta)}{\sqrt{z-\zeta}}G(z,\zeta)(\overline{G}-G(z))d\zeta\\
+i\sigma\frac{kax_0}{2}g(z)\int\limits_0^z\exp[i\Phi(z,\zeta)]\frac{u(x_0,\zeta)}{\sqrt{z-\zeta}}d\zeta,
\label{1r}
\end{multline}
where
\begin{align}
\Phi(z,\zeta)&=\frac{kax_0}{2}G(z,\zeta)-\frac{ka^2}{8}(z-\zeta)(\overline{G^2}-\overline{G}^2),\label{1s}\\
G(z,\zeta)&=\int\limits_\zeta^z g(\mu)d\mu,\quad \overline{G^2}=\frac{1}{z-\zeta}\int\limits_\zeta^z G^2(\mu)d\mu.\label{1t}
\end{align}
The obtained expressions can in principle be discretized for the use in a FD scheme -- cf.\cite{WM_Baskakov_1991, JCompPhys_Yevick_2001}.

Let us consider the case with constant curvature when $g(z)=1$. In this case $G(z)=z$, $\overline{G}=(z+\zeta)/2$, $\overline{G^2}=(z^2+z\zeta+\zeta^2)/3$ and $\overline{G^2}-\overline{G}^2=(z-\zeta)^2/12$. After substituting these expressions into (\ref{1q}) and (\ref{1r}) we obtain the following boundary conditions:

\begin{multline}
u(x_0,z)=\mp\frac{1}{\pi\sigma}\int\limits_0^z\exp[i\Phi(z-\zeta)]\frac{u'_x(x_0, \zeta)}{\sqrt{z-\zeta}}d\zeta\\
\pm i\frac{ka}{4\pi\sigma}\int\limits_0^z\exp[i\Phi(z-\zeta)]\sqrt{z-\zeta}u(x_0,\zeta)d\zeta,
\label{2a}
\end{multline}
\begin{multline}
u'_x(x_0,z)\mp\sigma\frac{a}{8}\int\limits_0^z\exp[i\Phi(z-\zeta)]\sqrt{z-\zeta}u'_x(x_0,\zeta)d\zeta=\\
\mp\sigma\frac{\partial}{\partial z}\int\limits_0^z\exp[i\Phi(z-\zeta)]\frac{u(x_0,\zeta)}{\sqrt{z-\zeta}}d\zeta\\
\mp i\sigma\frac{ka^2}{16}\int\limits_0^z\exp[i\Phi(z-\zeta)](z-\zeta)^{3/2}u(x_0, \zeta)d\zeta\\
\pm i\sigma\frac{kax_0}{2}\int\limits_0^z\exp[i\Phi(z-\zeta)]\frac{u(x_0,\zeta)}{\sqrt{z-\zeta}}d\zeta,
\label{2b}
\end{multline}
where
\begin{equation}
\Phi(t)=\frac{kax_0}{2}t-\frac{ka^2}{96}t^3,
\end{equation}
In (\ref{2a}) and (\ref{2b}) the upper sign corresponds to the upper boundary of the computational domain at $x_0$ and the lower sign to the lower boundary at $-x_0$. Eq. (\ref{2a})--(\ref{2b}) are convolutions. Let us assume now that $a=0$. Then the ordinary BPP conditions are obtained from (\ref{2a})--(\ref{2b}):
\begin{align}
u(x_0,z)&=\mp\frac{1}{\pi\sigma}\int\limits_0^z\frac{u'_x(x_0, \zeta)}{\sqrt{z-\zeta}}d\zeta, \label{2c}\\
u'_x(x_0,z)&=\mp\sigma\frac{\partial}{\partial z}\int\limits_0^z\frac{u(x_0,\zeta)}{\sqrt{z-\zeta}}d\zeta. \label{2d}
\end{align}
For a nonzero constant curvature case ($a\ne0$, $g(z)=1$) it is possible to obtain  from (\ref{2a}) and (\ref{2b}) the classical boundary conditions expressed through Airy functions \cite{RadSci_Popov_1996,ehrhardt2004solutions}. This can be done, for instance, by applying the Laplace transform by $z$ to convolution (\ref{2a}). (For details see \ref{ap2}.)

\section{TBC for the quadratic potential}
Let us consider the following PWE which is valid outside and at the boundary of our computational domain defined again as $|x|<x_0$
\begin{equation}
2ik\frac{\partial u}{\partial z}+\frac{\partial^2u}{\partial x^2}+k^2bx^2u=0,
\label{3a}
\end{equation}
where the parameter $b$ has been defined in (\ref{0c}) and is assumed to be non-negative. Eq. (\ref{3a}) can be transformed by introducing a new variable $\varphi(x,z)$:
\begin{equation}
u(x,z)=\varphi(x,z)\exp\left[\frac{\sqrt{b}}{2}(ikx^2-z)\right],
\label{3b}
\end{equation}
By substituting (\ref{3b}) into (\ref{3a}) one obtains the equivalent equation for the introduced function $\varphi$:
\begin{equation}
2ik\frac{\partial \varphi}{\partial z}+\frac{\partial^2\varphi}{\partial x^2}+2ik\sqrt{b}(x+x_0)\frac{\partial \varphi}{\partial x}=0,
\label{3d}
\end{equation}
where we inserted parameter $x_0$ to account for the fact the coordinate origin can be chosen arbitrary. The advantage of Eq. (\ref{3d}) is the linear dependence of one of its coefficients on coordinate $x$, which allows us, for instance, to obtain a TBC for (\ref{3d}) by applying the Laplace transform by variable $x$ and reducing it to the first order partial differential equation. Thus the quadratic case is slightly complicated than the linear case considered in the previous section.

It can be shown that the Laplace transform by $x$ (see Eq. (\ref{1e})) of Eq. (\ref{3d}) takes the following form
\begin{multline}
2ik\frac{\partial F}{\partial z}+2ik\sqrt{b}p\frac{\partial F}{\partial p}=\\
-(p^2+2ik\sqrt{b}(1+px_0))F+\varphi'(0, z)+(p+2ik\sqrt{b}x_0)\varphi(0, z)=0,
\label{3e}
\end{multline}
which is the linear first order differential equation with partial derivatives as was mentioned above. It can be solved by standard methods, which can be found, for instance, in \cite{miersemann2014partial}. The solution satisfying the initial condition $F(p,0)=0$ is
\begin{multline}
F=\frac{F_0(p)}{2ik\sqrt{b}}\int\limits_{p\exp(-\sqrt{b}z)}^p\frac{1}{F_0(p')}\times\\
\left[\frac{1}{p'}\varphi'\left(0, z-\frac{1}{\sqrt{b}}\ln\frac{p}{p'}\right)+\left(1+\frac{2ik\sqrt{b}x_0}{p'}\right)\varphi\left(0, z-\frac{1}{\sqrt{b}}\ln\frac{p}{p'}\right)\right]\,dp',
\label{3f}
\end{multline}
where
$$
F_0(p, z)=\frac{1}{p}\exp\left(-\frac{p^2}{4ik\sqrt{b}}-px_0\right).
$$
Introducing a new variable as $\mu=(\ln p/p')/\sqrt{b}$ and substituting it in $F_0$, the integral in (\ref{3f}) can be re-written as
\begin{multline}
F=\\
-\frac{i}{2k}\int\limits_0^z\exp\left(-\sqrt{b}\mu\right)\exp\left[-\frac{p^2}{4ik\sqrt{b}}\left(1-\exp(-2\sqrt{b}\mu)\right)-px_0\left(1-\exp(-\sqrt{b}\mu)\right)\right]\times\\
\left(g'_x(0,z-\mu)+(p\exp(-\sqrt{b}\mu)+2ik\sqrt{b}x_0)g(0,z-\mu)\right)\,d\mu,
\label{3g}
\end{multline}
Now applying the reverse Laplace transform (\ref{1h}), taking into account (\ref{1i})--(\ref{1l}) and again setting $c=0$ it is possible to show that the following Newman-to-Dirichlet mapping is obtained
\begin{multline}
\varphi(0,z)=-\frac{\sqrt{2\sqrt{b}}}{\pi\sigma}\int\limits_0^z\frac{\exp[ik\sqrt{b}x_0^2\tanh (\sqrt{b}(z-\mu)/2)]}{\sqrt{\exp[2\sqrt{b}(z-\mu)]-1}}\varphi'_x(0,\mu)d\mu-\\
2ik\sqrt{b}x_0\frac{\sqrt{2\sqrt{b}}}{\pi\sigma}\int\limits_0^z\frac{\exp[ik\sqrt{b}x_0^2\tanh (\sqrt{b}(z-\mu)/2)]}{\sqrt{\exp[2\sqrt{b}(z-\mu)]-1}}\frac{\varphi(0,\mu)}{1+\exp[-\sqrt{b}(z-\mu)]}d\mu.
\label{3h}
\end{multline}
The reverse mapping similar to expression in (\ref{1p}) can also be derived. Finally, using definition (\ref{3b}) we can obtain for amplitude $u$
\begin{multline}
u(0,z)=\mp\frac{\sqrt{2\sqrt{b}}}{\pi\sigma}\int\limits_0^z\frac{\exp[\Phi(z-\mu)]}{\sqrt{\exp[2\sqrt{b}(z-\mu)]-1}}u'_x(0,\mu)d\mu\mp\\
ik\sqrt{b}x_0\frac{\sqrt{2\sqrt{b}}}{\pi\sigma}\int\limits_0^z\frac{\exp[\Phi(z-\mu)]}{\sqrt{\exp[2\sqrt{b}(z-\mu)]-1}}\tanh(\sqrt{b}(z-\mu)/2)u(0,\mu)d\mu,
\label{3i}
\end{multline}
where
\begin{equation}
\Phi(s)=-\sqrt{b}s/2+ik\sqrt{b}x_0^2\tanh(\sqrt{b}s/2).
\label{3j}
\end{equation}
In (\ref{3i}) the upper sign corresponds to the upper boundary of the computational domain at $x_0$ and the lower sign to the lower boundary at $-x_0$. It is clear that when $b\to0$, expression (\ref{3i}) goes to the BPP condition (\ref{2c}). Note that the phase like value (\ref{3j}) has a growing negative real part and its imaginary part goes to a constant when $\mu\to0$, which corresponds to the points furthest from the current position at $z$. This fact means that these remote points have little or no influence on the integrals in (\ref{3i}), which is not surprising for a strongly leaking waveguide (or a repulsive potential in case of the 1D SE) corresponding to the case when $b>0$. In such a waveguide or potential the boundary functions as a sink for the waves incident on it. On contrary, when $b<0$, expression (\ref{3i}) becomes
\begin{multline}
u(0,z)=\mp\frac{\sqrt{2i\sqrt{|b|}}}{\pi\sigma}\int\limits_0^z\frac{\exp[\Phi(z-\mu)]}{\sqrt{\exp[2i\sqrt{|b|}(z-\mu)]-1}}u'_x(0,\mu)d\mu\pm\\
ik\sqrt{|b|}x_0\frac{\sqrt{2i\sqrt{|b|}}}{\pi\sigma}\int\limits_0^z\frac{\exp[\Phi(z-\mu)]}{\sqrt{\exp[2i\sqrt{|b|}(z-\mu)]-1}}\tan(\sqrt{|b|}(z-\mu)/2)u(0,\mu)d\mu,
\label{3k}
\end{multline}
where
\begin{equation}
\Phi(s)=-i\sqrt{|b|}s/2-ik\sqrt{|b|}x_0^2\tan(\sqrt{|b|}s/2).
\label{3l}
\end{equation}
Now the phase like value (\ref{3l}) is purely imaginary and, moreover, it has a periodic term with singularities. This fact means that all points from $0$ to $z$ has strong influence on the integrals in (\ref{3k}), which can be explained by reflection of waves from the boundary in case of a confining waveguide or an attractive potential. In the more general case when $b$ is a complex number, the behavior of integrals is intermediate between these two extreme cases. 

\section{Discussion and conclusion}
In this work we derived two exact transparent boundary boundary conditions for the 2D PWE (or for the equivalent 1D SE). The first of them is for a linear dependence of the permittivity (or potential in case of 1D SE) on transversal coordinate $x$ (see Eq. (\ref{0b})) with the coefficient also dependent on the longitudinal coordinate $z$. The second TBC is for a quadratic dependence on $x$ with constant coefficients (see Eq. (\ref{0c})). As is usual for the TBCs they are non-local by the longitudinal coordinate (or time for the 1D SE). The integral kernels of both TBCs contain only elementary functions and relate the derivative of the field amplitude by the transversal coordinate at the boundary of the computational domain to the amplitude itself. This fact favorably distinguishes them from the known transparent boundary conditions, where the kernels are defined as some complex integrals over expressions containing special functions. This makes such conditions often difficult to apply in practice.

For the practical application of both TBCs they must be discretized to be used with a specific FD scheme. For the popular Crank--Nicholson FD scheme \cite{WM_Baskakov_1991, antoine2003unconditionally} such a discretization may be done by using linear interpolation of slowly changing field amplitudes at the boundaries of the computational domain. The main problem remains how to develop a stable and accurate numerical integration method for the integral kernels in (\ref{1q})--(\ref{1r}) and (\ref{3i})--(\ref{3l}) that contain weak singularities and fast oscillating functions. We are planning to discuss this issue elsewhere.  

If the problems of discretization are overcome, the TBCs derived in the present paper may significantly simplify the numerical solution of the 2D PWE (or the equivalent 1D SE) by FD schemes in cases when the dielectric permittivity or potential have a linear or quadratic dependence on the transversal coordinate outside the computational domain. They may find applications also in 3D (2D and 3D SE) case if the FD scheme for a higher dimension equation is reduced to FD schemes of lower dimensions at the boundary of the computational domain as it was done, for instance, in \cite{JOSA_A_Feshchenko_2011, feshchenko2013exact}. 

\section{Acknowledgment}
The authors would like to thank A.V. Vinogradov for the fruitful discussions about the parabolic and Schr\"odinger equations and their applications.

\appendix
\section{Integral calculation}
\label{ap1}
In order to derive equality (\ref{1l}) let us find the following limit
\begin{multline}
\lim_{x\to0}\int\limits_{-i\infty}^{i\infty}\exp(i\beta w^2)\exp(xw)wdw=\\
\lim_{x\to0}\frac{\partial}{\partial x}\int\limits_{-i\infty}^{i\infty}\exp(i\beta(w+x/(2i\beta))^2+ix^2/(4\beta))dw=\\
\lim_{x\to0}\frac{ix}{2\beta}\sqrt{\frac{\pi i}{\beta}}\exp(ix^2/(4\beta))=
\left\{
\begin{array}{l}
0,\;\mbox{if}\;\beta\ne0\\
\mbox{not defined},\;\mbox{if}\;\beta=0
\end{array}
\right..
\label{ap1a}
\end{multline}
Taking into account that
\begin{equation}
\int\limits_{-\infty}^\infty\frac{ix}{2\beta}\sqrt{\frac{\pi i}{\beta}}\exp(ix^2/(4\beta))\;d\beta=-2\pi,
\label{ap1b}
\end{equation}
we conclude that 

\begin{equation}
\lim_{x\to0}\int\limits_{-i\infty}^{i\infty}\exp(i\beta w^2)\exp(xw)wdw=-2\pi\delta(\beta),
\label{ap1c}
\end{equation}

Let us derive equality (\ref{1m}). We have that
\begin{multline}
\lim_{x\to0}\int\limits_{-i\infty}^{i\infty}\int\limits_0^z f(\zeta)\exp(i(z-\zeta)w^2/2k)\exp(xw)w^2d\zeta dw=\\
\lim_{x\to0}\frac{\partial^2}{\partial x^2}\int\limits_{-i\infty}^{i\infty}\int\limits_0^z f(\zeta)\exp(i(z-\zeta)w^2/2k)\exp(xw)d\zeta dw=\\
\lim_{x\to0}\int\limits_0^z\frac{\partial^2}{\partial x^2}\sqrt{\frac{2k\pi}{-i(z-\zeta)}}\exp(ikx^2/(z-\zeta)/2)d\zeta=\\
-i^{3/2}(2k)^{3/2}\sqrt{\pi}\lim_{x\to0}\int\limits_0^z\frac{\partial}{\partial z}\frac{1}{\sqrt{z-\zeta}}\exp(ix^2/(z-\zeta)/2)d\zeta=\\
(2k)^{3/2}\sqrt{\frac{\pi}{i}}\left\{\lim_{x\to0}\frac{\partial}{\partial z}\int\limits_0^z\frac{1}{\sqrt{z-\zeta}}\exp(ix^2/(z-\zeta)/2)d\zeta-\right.\\
\left.\phantom{\int\limits_0^z}\lim_{x\to0}\lim_{\zeta\to z}\frac{1}{\sqrt{z-\zeta}}\exp(ix^2/(z-\zeta)/2)\right\}.
\label{ap1d}
\end{multline}
In the last expression the limit of the first term can be found by substituting $x=0$. The second term is zero because it contains an oscillating exponent at its limit. The final expression is
\begin{multline}
\lim_{x\to0}\int\limits_{-i\infty}^{i\infty}\int\limits_0^z f(\zeta)\exp(i(z-\zeta)w^2/2k)\exp(xw)w^2d\zeta dw\\
=\sqrt{\frac{\pi}{i}}(2k)^{3/2}\frac{\partial}{\partial z}\int\limits_0^z\frac{f(\zeta)}{\sqrt{z-\zeta}}d\zeta
\label{ap1e}
\end{multline}

\section{Derivation of the classical condition with Airy functions}
\label{ap2}
Applying the Laplace transform by $z$ to condition (\ref{2a}) we obtain that
\begin{equation}
\label{ap2a}
F'_x(x,p)=-\pi\sigma\frac{1+\frac{ia}{4k\pi\sigma}\frac{\partial I_1(p')}{\partial p'}}{I_1(p')}F(x,p),
\end{equation}
where
$$
F(x,p)=\int\limits_0^\infty u(x,z)\exp(-pz)dz, \quad p'=p-\frac{iax}{2k},
$$
and integral $I_1$ is defined as
\begin{equation}
\label{ap2a1}
I_1=\int\limits_0^\infty\exp\left[-p'\zeta-\frac{ia^2}{96k^3}\zeta^3\right]\frac{d\zeta}{\sqrt{\zeta}}.
\end{equation}
Now applying the reverse Laplace transform to (\ref{ap2a}) we obtain that
\begin{equation}
\label{ap2b}
\frac{\partial u(x,z)}{\partial x}=\frac{\partial}{\partial z}\int\limits_0^z K(z-\zeta)u(x,\zeta)d\zeta,
\end{equation}
where the kernel $K$ can be written as
\begin{align}
K(z)&=-\frac{a^{1/3}}{2\pi i}\int\limits_{-\infty+ic}^{\infty+ic}\frac{\exp[i\xi t]}{t}\frac{w_1'(t-t_0)}{w_1(t-t_0)}dt,\quad a>0,\label{ap2c}\\
K(z)&=-\frac{|a|^{1/3}}{2\pi i}\int\limits_{-\infty+ic}^{\infty+ic}\frac{\exp[i\xi t]}{t}\frac{\mathrm{Ai}'(t-t_0)}{\mathrm{Ai}(t-t_0)}dt,\quad a<0.\label{ap2d}
\end{align}
Here
$$
\xi=z\frac{|a|^{2/3}}{2k},\quad t_0=a^{1/3}x,\quad w_1(t)=\mathrm{Bi}(t)+i\mathrm{Ai}(t),
$$
and $\mathrm{Ai}$ and $\mathrm{Bi}$ are Airy functions. To obtain formulas (\ref{ap2c}) and (\ref{ap2d}) we used the following equalities:
\begin{align}
I_1&=\frac{2^{5/6}\sqrt{k}}{i^{1/6}|a|^{1/3}}M(w),\quad w=p'\frac{k 2^{5/3}}{i^{1/3}a^{2/3}},\label{ap2e}\\
M(w)&=\int\limits_0^\infty\frac{\exp[-wt-t^3/3]}{\sqrt{t}}dt=\frac{\pi^{3/2}}{2^{1/3}}\left[\mathrm{Ai}^2(-w/2^{2/3})+\mathrm{Bi}^2(-w/2^{2/3})\right]\label{ap2f}\\
&=2^{5/3}\pi^{3/2}\mathrm{Ai}(w i^{2/3}/2^{2/3})\mathrm{Ai}(w i^{-2/3}/2^{2/3}),\notag\\
\mathrm{Ai}(w i^{2/3}&/2^{2/3})\mathrm{Ai}'(w i^{-2/3}/2^{2/3})-i^{4/3}\mathrm{Ai}'(w i^{2/3}/2^{2/3})\mathrm{Ai}(w i^{-2/3}/2^{2/3})=\frac{i^{5/3}}{2\pi}.\label{ap2g}
\end{align}
Expressions (\ref{ap2c}) and (\ref{ap2d}) coincide with those reported in \cite{RadSci_Popov_1996}.

\section*{\refname}

\end{document}